\documentclass
[prl,superscriptaddress,nofootinbib,twocolumn,
showpacs,preprintnumbers,amsmath,amssymb,a4wide]
{revtex4-1}

\usepackage{graphicx}
\usepackage{hyperref}
\usepackage{subfigure}
\usepackage{bm}
\usepackage{epsfig}
\usepackage{amsmath, amsthm, amsfonts}
\usepackage{color}
\usepackage{enumitem}
\usepackage{array}

\newcommand{\be}{\begin{equation}}
\newcommand{\ee}{\end{equation}}
\newcommand{\ba}{\begin{eqnarray}}
\newcommand{\ea}{\end{eqnarray}}
\newcommand{\ban}{\begin{eqnarray*}}
\newcommand{\ean}{\end{eqnarray*}}

\newcommand{\ket}[1]{\mbox{$ | #1 \rangle $}}

\begin{document}

\title{Classical simulation of entanglement swapping with bounded communication}

\author{Cyril Branciard}
\affiliation{School of Mathematics and Physics, The University of Queensland, St Lucia, QLD 4072, Australia}
\author{Nicolas Brunner}
\affiliation{H.H. Wills Physics Laboratory, University of Bristol, Bristol, BS8 1TL, United Kingdom}
\author{Harry Buhrman}
\affiliation{Centrum Wiskunde \& Informatica, and University of Amsterdam, Science Park 123, 1098 XG Amsterdam, The Netherlands}
\author{Richard Cleve}
\affiliation{Institute for Quantum Computing and School of Computer Science, University of Waterloo, 200 University Avenue West, Waterloo, Ontario, Canada N2L 3G1}
\affiliation{Perimeter Institute for Theoretical Physics, 31 Caroline Street North, Waterloo, Ontario, Canada N2L 2Y5}
\author{Nicolas Gisin}
\affiliation{Group of Applied Physics, University of Geneva, Chemin de Pinchat 22, CH-1211 Geneva 4, Switzerland}
\author{Samuel Portmann}
\affiliation{Group of Applied Physics, University of Geneva, Chemin de Pinchat 22, CH-1211 Geneva 4, Switzerland}
\author{Denis Rosset}
\affiliation{Group of Applied Physics, University of Geneva, Chemin de Pinchat 22, CH-1211 Geneva 4, Switzerland}
\author{Mario Szegedy}
\affiliation{Department of Computer Science,
Rutgers, the State University of NJ,
110 Frelinghuysen Road,
Piscataway, NJ 08854-8019
USA}

\date{\today}

\begin{abstract}
Entanglement appears under two different forms in quantum theory, namely as a property of states of joint systems and as a property of measurement eigenstates in joint measurements. By combining these two aspects of entanglement, it is possible to generate nonlocality between particles that never interacted, using the protocol of entanglement swapping. We show that even in the more constraining bilocal scenario where distant sources of particles are assumed to be independent, i.e.~to share no prior randomness, this process can be simulated classically with bounded communication, using only 9 bits in total. Our result thus provides an upper bound on the nonlocality of the process of entanglement swapping.
\end{abstract}


\maketitle

By performing suitably chosen local measurements on an entangled quantum state, distant observers can establish nonlocal correlations, as witnessed by the violation of a Bell inequality~\cite{bell64}. This means that quantum statistics cannot be simulated by classically correlated systems, unless some classical communication is added to the model. Although experiments give strong evidence that nature does not use classical communication to establish correlations~\cite{salart08}, it is nevertheless interesting from a fundamental perspective to ask how much communication is required to reproduce quantum correlations. Generally referred to as classical simulation of entanglement, this provides a natural approach to the problem of quantifying quantum nonlocality.

Nonlocality is a fundamental aspect of quantum mechanics, hence quantifying it is much desirable. Besides being one of the most striking and counterintuitive feature of the theory, it is also a powerful resource, allowing for instance for the reduction of communication complexity~\cite{buhrman10}, as well as for information processing in the 'device-independent' setting~\cite{ekert91,barrett05,acin07,pironio10,colbeck11}, where one wants to achieve an information task and prove its security without any assumption on the devices used in the protocol.

Several works~\cite{maudlin92,brassard99,gisin99,steiner00} underwent the task of estimating how much communication is needed to simulate the correlations of a maximally entangled state of two qubits under all possible projective measurements. This research culminated in 2003, when Toner and Bacon~\cite{toner03} showed that one bit of communication is enough. Importantly this single bit of communication is not an average value, but represents the exact amount that is to be used at each round. Thus the model is said to have bounded communication. The communication costs of other states have been explored as well~\cite{pironio03,degorre07}. Notably, Regev and Toner~\cite{regev07} have shown that the correlations obtainable from dichotomic measurements on any bipartite entangled state can be simulated with only two bits of communication, which are proven to be necessary~\cite{vertesi09}. Note however that their protocol does not reproduce the correct marginal distributions, and that simulating more general measurements is much more costly in terms of communication~\cite{brassard99,buhrman10}. The simulation of multipartite entanglement also attracted some attention~\cite{tessier05,barrett07,broadbent09,bancal10}, and two of the authors~\cite{branciard11} recently showed that the correlations of equatorial measurements on a tripartite GHZ state can be simulated with 3 bits of communication, thus reproducing in particular the Mermin-GHZ paradox, which is arguably the strongest demonstration of the nonlocality of this state~\cite{greenberger89,mermin90}.

Quantum mechanics allows not only for entangled states of distant systems, but also for entangled measurements. In such a measurement the initial state is arbitrary---it could be entangled or not---but the final state is entangled, that is the eigenstates of the operator that represents such a measurement are entangled. This second aspect of entanglement is in itself independent of nonlocality---although it leads to nonlocality when combined with entangled states~\cite{grudka08}. It demonstrates another nonclassical feature of entanglement, which is, loosely speaking, the possibility to ask two (or more) quantum systems questions about their relations without gaining any information about the individual properties of each subsystem~\cite{gisin06}.

\begin{figure}
\centering
\epsfig{file=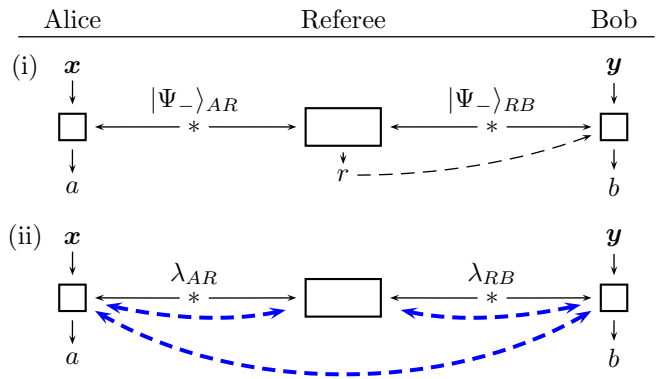}
\caption{(i) The scenario of entanglement swapping with two fully independent sources of $\ket{\Psi_-}$ states. The Referee sends to Bob the result $r$ of his Bell state measurement. Upon receiving these two bits of communication, Bob can apply the adequate unitary operation to his qubit such that Alice and Bob finally share a singlet state. (ii) The classical simulation of entanglement swapping in the bilocal scenario, i.e.~with two uncorrelated sources of shared random variables $\lambda_{AR}$ and $\lambda_{RB}$. The three parties exchange messages symbolized by the thick dashed arrows. Here we present a simulation protocol using 9 bits of communication in total.}
\label{fig.swap}
\end{figure}

It is therefore a natural question to ask whether or not the classical simulation of protocols involving both entangled states and entangled measurements is possible with bounded communication, and how much communication is required. Here, we investigate this question for the scenario of entanglement swapping~\cite{zukowski93}, where quantum particles that never interacted become nonlocally correlated after their twins underwent a joint measurement. We work in the scenario of \emph{bilocality}~\cite{branciard10, branciard112}, in which the shared randomness we will use in the simulation protocol is assumed to originate from two independent sources (see figure~\ref{fig.swap}). Since entanglement swapping can be achieved with fully uncorrelated quantum sources, even experimentally~\cite{halder07,kaltenbaek09}, it is indeed natural to impose an equivalent constraint on the simulation model.


Importantly the constraint of bilocality makes the simulation of entanglement swapping a challenging problem. The goal is basically to generate singlet nonlocal correlations between two parties which are initially fully uncorrelated. At first sight this may even appear to be impossible with finite communication, in the light of a result by Massar et al.~\cite{massar01}, showing that simulating singlet correlations requires either infinite shared randomness or infinite communication. Here however, we will show that the process of entanglement swapping can be simulated with 9 bits, by presenting an explicit protocol. Since the communication cost of the cheapest protocol can be considered as a measure of its nonlocality, our result provides an upper bound on the nonlocality of entanglement swapping.

\emph{The entanglement swapping process.} We consider three distant parties, Alice~($A$), Bob~($B$) and a Referee~($R$). $R$ shares two maximally entangled qubit pairs in the state $\ket{\Psi_-} = \frac{1}{\sqrt{2}}(\ket{01} - \ket{10})$ with $A$ and $B$, respectively. The first pair is produced by a source located between $A$ and $R$, the second one is produced by an independent source located between $B$ and $R$, see Fig.~\ref{fig.swap}~(i). Accordingly, Alice and Bob are initially uncorrelated. By performing a Bell state measurement, i.e.~a two-qubit joint measurement which features four maximally entangled Bell states as eigenstates, $R$ projects $A$ and $B$'s particles onto one of the Bell states. $R$ can thus 'swap' entanglement to $A$ and $B$. This protocol is essentially identical to the celebrated quantum teleportation protocol~\cite{bennett93}; entanglement swapping is basically a {\it teleportation of entanglement}.


To complete the protocol, the Referee needs to communicate the result of his measurement $r$---by sending two bits of classical communication---to (say) Bob, who can then apply a suitable unitary local transformation to his qubit to finally share a definite Bell state with Alice, say $\ket{\Psi_-}$. Upon receiving measurement settings $\bm{x}$ and $\bm{y}$ (represented by unit vectors on the Bloch sphere) and performing the corresponding projective measurements on their respective qubit, Alice and Bob obtain binary measurement outcomes $a=\pm 1$ and $b=\pm1$ respectively. These outcomes exhibit nonlocal correlations of the form
\ba
E(\bm{x},\bm{y}) &=& P(a=b|\bm{x},\bm{y}) - P(a \neq b|\bm{x},\bm{y}) = - \bm{x} \cdot \bm{y}, \quad \ \label{corr}
\ea
with random marginals.


A classical simulation of this protocol in the bilocal scenario would then amount to the following (see Fig.~\ref{fig.swap}~(ii)): Alice and Bob receive as inputs the measurement directions $\bm{x}$ and $\bm{y}$, independently of the random variables $\lambda_{AR}$ and $\lambda_{RB}$ they may each share (independently) with $R$. After a finite amount of information exchange between the three parties, Alice and Bob produce their outputs which, after averaging over the shared random variables, must be correlated as in~\eqref{corr}~\cite{footnote_simul_BSM}.

At this point, it is instructive to recall the result of Massar \textit{et al.}~\cite{massar01}, which implies that correlations of the form~\eqref{corr} are impossible to simulate with finite communication when $A$ and $B$ only have finite \textit{shared} randomness (in the sense that their shared randomness could be established with bounded communication)---even if they may each have an additional source of infinite randomness, independent for $A$ and $B$.
The present scenario may at first glance appear similar: initially, $A$ and $B$ each only have access to independent sources of infinite randomness, $\lambda_{AR}$ and $\lambda_{RB}$, respectively. After any finite exchange of communication between the parties (including $R$), they could only share some finite amount of randomness.
Intuition suggests that the result of~\cite{massar01} implies that, from such randomness, 
correlations of the form~\eqref{corr} cannot be simulated with finite communication, and entanglement swapping is impossible to simulate in a bilocal manner~\cite{footnote_erroneous_preprint}.
However, the present scenario differs from that of \cite{massar01} in that the third party, $R$, can share an infinite amount of randomness with $A$ and $B$ independently---by having access to both $\lambda_{AR}$ and $\lambda_{RB}$--- and can exchange some (finite number of) bits with $A$ and $B$.
This seemingly subtle difference dramatically changes the situation: despite the above heuristic reasoning, entanglement swapping can be simulated with finite communication in a bilocal manner, as we now show.

\emph{Simulation protocol.} We start by deriving a protocol for the simulation of equatorial measurements $\bm{x}=(\cos{\phi_A},\sin{\phi_A},0)$ and $\bm{y}=(\cos{\phi_B},\sin{\phi_B},0)$. In this case, the correlation~\eqref{corr} is equivalent to
\ba
P(a{=}b|\phi_A,\phi_B) = \frac{1{-}\cos(\phi_A{-}\phi_B)}{2} = \sin\!\big(\frac{\phi_A{-}\phi_B}{2}\big)^2. \quad \label{corr_equat}
\ea
Note in particular that if $\phi_A=\phi_B$, then $P(a=b|\phi_A=\phi_B) = 0$, as expected from the anticorrelations of the singlet state.

Let us divide the equator of the Bloch sphere into 2$m$ sectors of the same size, and assume that Alice and the Referee (resp.~the Referee and Bob) share a random variable $\lambda_{AR}$ (resp. $\lambda_{RB}$), with $\lambda_{AR}$ and $\lambda_{RB}$ uniformly distributed in $[0,\frac{\pi}{m}]$. Let Alice tell Bob in which sector her setting (modulo $\pi$) lies, and send him the additional bit $c_A = [(\phi_A \mod \frac{\pi}{m}) < \lambda_{AR}]$ (we use the Iverson bracket, with $[{\cal E}] = 1$ if the expression ${\cal E}$ is true, $[{\cal E}] = 0$ otherwise); let the Referee send to Bob the bit $c_R = [\lambda_{AR} < \lambda_{RB}]$; and let Bob calculate $c_B = [\lambda_{RB} < (\phi_B \mod \frac{\pi}{m})]$.

To warm up and gain some intuition, assume that Alice and Bob's settings both lie in $[0,\frac{\pi}{m}]$, and let Alice simply output $a=+1$. If $c_A = c_R = c_B$, then Bob knows that (with probability 1) $\phi_A \neq \phi_B$; let him then also (for now) output $b = a = +1$. On the other hand, if $c_A \neq c_R$ or $c_R \neq c_B$, it could be the case that $\phi_A = \phi_B$; to ensure that $P(a=b|\phi_A=\phi_B) = 0$, Bob must output $b = -a = -1$. The probability that Alice and Bob give the same output is then (assuming $\phi_A < \phi_B$; the case $\phi_A \geq \phi_B$ being similar)
\ba
P(a{=}b|\phi_A < \phi_B) &=& P(c_A=c_B=c_R | \phi_A < \phi_B) \nonumber \\
&& \hspace{-2.5cm} = P(\phi_A < \lambda_{AR} < \lambda_{RB} < \phi_B | \phi_A < \phi_B) \nonumber \\
&& \hspace{-2.5cm} = \int_{\phi_A}^{\phi_B} \frac{m}{\pi} d\lambda_{AR} \int_{\lambda_{AR}}^{\phi_B} \frac{m}{\pi} d\lambda_{RB} = \frac{m^2}{2\pi^2} (\phi_A{-}\phi_B)^2. \ \quad \label{eq_Prot0}
\ea
If $m \geq \frac{\pi}{\sqrt{2}}$, this probability is too large (for any $\phi_A, \phi_B \in [0,\frac{\pi}{m}]$), compared to the desired value of $\sin\!\big(\frac{\phi_A{-}\phi_B}{2}\big)^2$~\eqref{corr_equat}. Now, one can actually make use of some freedom in the case $c_A = c_R = c_B$, and ask Bob to output $b = a = +1$ only with some probability $\wp$ (and to output $b = -a = -1$ with probability $1-\wp$). By letting $\wp$ depend on $\phi_B$ and $\lambda_{RB}$, and take the particular form $\wp(\phi_B{-}\lambda_{RB}) = \frac{\pi^2}{2m^2} \cos(\phi_B{-}\lambda_{RB})$, we obtain, as desired,
\ba
P(a{=}b|\phi_A,\phi_B) &=& \frac{m^2}{\pi^2} \int_{\phi_A}^{\phi_B} \!\!\! d\lambda_{AR} \int_{\lambda_{AR}}^{\phi_B} \!\!\! d\lambda_{RB} \ \wp(\phi_B{-}\lambda_{RB}) \nonumber \\
&=& \sin\!\big(\frac{\phi_A{-}\phi_B}{2}\big)^2. \ \quad
\ea

In order now to extend this protocol to all possible equatorial measurements $\phi_A, \phi_B \in [0,2\pi]$, one needs to find adequate strategies for Bob (i.e.~adequate functions $\wp(\phi_B{-}\lambda_{RB}) \in [0,1]$) for all possible sectors where Alice and Bob's settings may lie, and for all possible values of $c_A, c_R$ and $c_B$. We were able to find such strategies for the case $m=4$, see Table~\ref{table_ps}. This leads us to define the following protocol:


\newcolumntype{P}[1]{>{\centering $}p{#1}<{$}}
\renewcommand{\arraystretch}{1.4}

\begin{table*}

\begin{tabular}{P{1.5cm}||P{3.6cm}|P{3.6cm}|P{3.6cm}|P{3.6cm}|}
c_A\,c_R\,c_B & j_B=0, \ \gamma \in [{-}\frac{\pi}{4},\frac{\pi}{4}] & j_B=1, \ \gamma \in [0,\frac{\pi}{2}] & j_B=2, \ \gamma \in [\frac{\pi}{4},\frac{3\pi}{4}] & j_B=3, \ \gamma \in [\frac{\pi}{2},\pi]
 \tabularnewline
 \hline
 \hline
0\ 0\ 0 & \frac{\pi^2}{32}\cos\gamma & 0 & \frac{1}{2}{+}\frac{\pi^2}{64}[1{-}(2{+}\sqrt{2})\sin\gamma] & 1+\frac{\pi^2}{32}[\cos\gamma{+}2\cos(\gamma{+}\frac{\pi}{4})]
 \tabularnewline
 \hline
0\ 0\ 1 & 0 & 0 & \frac{1}{2}{+}\frac{\pi^2}{64}[1{-}\sqrt{2}\sin\gamma] & 1+\frac{\pi^2}{32}\cos\gamma
 \tabularnewline
 \hline
0\ 1\ 0 & 0 & 0 & \frac{1}{2}{-}\frac{\pi^2}{64}[1{+}\sqrt{2}\sin(\gamma{+}\frac{\pi}{4})] & 1+\frac{\pi^2}{32}\cos(\gamma{+}\frac{\pi}{4})
 \tabularnewline
 \hline
0\ 1\ 1 & 0 & \frac{\pi^2}{32}\cos(\gamma{-}\frac{\pi}{4}) & \frac{1}{2}{-}\frac{\pi^2}{64}[1{-}\sqrt{2}\sin(\gamma{-}\frac{\pi}{4})] & 1
 \tabularnewline
 \hline
1\ 0\ 0 & 0 & 0 & \frac{1}{2}{+}\frac{\pi^2}{64}[1{-}\sqrt{2}\sin(\gamma{+}\frac{\pi}{4})] & 1+\frac{\pi^2}{32}\cos(\gamma{+}\frac{\pi}{4})
 \tabularnewline
 \hline
1\ 0\ 1 & 0 & \frac{\pi^2}{32}\cos(\gamma{-}\frac{\pi}{4}) & \frac{1}{2}{+}\frac{\pi^2}{64}[1{+}\sqrt{2}\sin(\gamma{-}\frac{\pi}{4})] & 1
 \tabularnewline
 \hline
1\ 1\ 0 & 0 & \frac{\pi^2}{32}\cos\gamma & \frac{1}{2}{-}\frac{\pi^2}{64}[1{-}\sqrt{2}\sin\gamma] & 1
 \tabularnewline
 \hline
1\ 1\ 1 & \frac{\pi^2}{32}\cos\gamma & \frac{\pi^2}{32}[\cos\gamma{+}2\cos(\gamma{-}\frac{\pi}{4})] & \frac{1}{2}{-}\frac{\pi^2}{64}[1{-}(2{+}\sqrt{2})\sin\gamma] & 1
 \tabularnewline
 \hline
\end{tabular}

\caption{Functions $\wp_{c_A c_R c_B}^{j_B}(\gamma)$, for all values of $j_B, c_A, c_R$ and $c_B$, defining the probability $\wp_{c_A c_R c_B}^{j_B}(\phi_B'{-}\lambda_{RB})$ that Bob outputs $b = \beta$ in Protocol~1. Note that in all cases, $\wp_{c_A c_R c_B}^{j_B}(\gamma) \in [0,1]$ for all possible values of $\gamma = \phi_B'{-}\lambda_{RB}$ in the given interval.}
\label{table_ps}

\end{table*}

\medskip

{\bf Protocol~1:} {\it Let Alice and the Referee (resp. the Referee and Bob) share a random variable $\lambda_{AR}$ (resp. $\lambda_{RB}$), with $\lambda_{AR}$ and $\lambda_{RB}$ uniformly distributed in $[0,\frac{\pi}{4}]$. After reception of Alice and Bob's measurement settings $\phi_A, \phi_B$, the three parties proceed as follows:

\begin{itemize}[leftmargin=*,topsep=2pt,itemsep=2pt,parsep=2pt]

\item Alice calculates $a = {\mathrm{sign}}(\sin \phi_A) = \pm 1$, $j_A = \lfloor \frac{4}{\pi}(\phi_A \mod \pi) \rfloor \in \{0,1,2,3\}$, 
$\phi_A' = (\phi_A{-}j_A\frac{\pi}{4} \mod \pi) \in [0,\frac{\pi}{4}[$ and $c_A = [\phi_A' < \lambda_{AR}] \in \{0,1\}$. She sends $j_A$ (2 bits) and $c_A$ (1 bit) to Bob, and outputs $a$.

\item The Referee calculates $c_R = [\lambda_{AR} < \lambda_{RB}] \in \{0,1\}$ and sends it to Bob (1 bit).

\item After reception of $j_A$, Bob calculates $\beta = {\mathrm{sign}}[\sin (\phi_B-j_A\frac{\pi}{4})]$ and $\phi_B' = (\phi_B-j_A\frac{\pi}{4} \mod \pi) \in [0,\pi[$. He determines the index $j_B = \lfloor \frac{4}{\pi}\phi_B' \rfloor \in \{0,1,2,3\}$ of the sector where his angle $\phi_B'$ lies, and the bit $c_B = [\lambda_{RB} < \phi_B'{-}j_B\frac{\pi}{4} ] \in \{0,1\}$. Depending on $j_B, c_A, c_R, c_B, \phi_B'$ and $\lambda_{RB}$, he outputs $b = \beta$ with probability $\wp_{c_A c_R c_B}^{j_B}(\phi_B'{-}\lambda_{RB})$, and $b = -\beta$ with probability $1-\wp_{c_A c_R c_B}^{j_B}(\phi_B'{-}\lambda_{RB})$, for the functions $\wp_{c_A c_R c_B}^{j_B}$ defined in Table~\ref{table_ps}.

\end{itemize}}

\medskip

As explicitly shown in the Appendix below, the above protocol gives the desired probability $P(a=b|\phi_A,\phi_B) = \sin(\frac{\phi_A-\phi_B}{2})^2$ for all possible equatorial measurements, using 4 bits of communication. It can then be extended in the following way to all measurement directions on the Bloch sphere, using a similar technique as in~\cite{brassard99}:

\medskip

{\bf Protocol~2:} {\it For measurement directions $\bm{x} =$ $(\sin{\theta_A}\cos{\phi_A}, \sin{\theta_A}\sin{\phi_A}, \cos{\theta_A})$ and $\bm{y} =$ $(\sin{\theta_B}\cos{\phi_B},$ $\sin{\theta_B}\sin{\phi_B}, \cos{\theta_B})$, the three parties

\begin{itemize}[leftmargin=*,topsep=2pt,itemsep=2pt,parsep=2pt]

\item run Protocol~1 with input angles $\phi_A$ and $\phi_B$; Alice and Bob obtain intermediate outputs $a_0$ and $b_0$.

\item run Protocol~1 a second time (using a new set of independent variables $\lambda_{AR}$, $\lambda_{RB}$), now with input angles $a_0 \theta_A$ and $-b_0 \theta_B$; Alice and Bob output the outcomes $a$ and $b$ of this second run of Protocol~1.

\end{itemize}}

\medskip

This second protocol now simulates the desired correlation $E(\bm{x},\bm{y}) = - \bm{x} \cdot \bm{y}$ for all possible projective measurements by Alice and Bob, with 8 bits of communication; for more details on the calculations, see the Appendix. Note that Protocols~1 and~2 do not simulate the correct marginals. In order to randomize the marginals, Alice can---at the very end of the protocol---generate a random bit and send it to Bob; depending on the value of this bit, they will both flip their outcomes or not. All in all, the entanglement swapping correlations can thus be simulated with 9 bits of communication.

\paragraph{Discussion.}

We thus have proved that remarkably, the entanglement swapping process can be simulated with bounded communication, even in a bilocal scenario where Alice and Bob are (as in the quantum case) completely uncorrelated before the protocol is run, and therefore do not have any prior shared randomness. Our protocol provides an upper bound on the nonlocality of entanglement swapping in terms of its communication cost. It is an open question whether fewer bits of communication are actually sufficient: it might indeed be possible to simulate equatorial measurements more efficiently than with Protocol~1, or to find a more direct protocol that does not treat separately the azimuth and zenith angles of the measurement settings, more in the spirit of the Toner-Bacon simulation protocol for singlet correlations~\cite{toner03}.

Next, it is natural to consider the simulation of multistage entanglement swapping, which is essential for long distance quantum communication. Now, $N$ referees ($R_1$, $R_2$, $\dots$, $R_N$) are placed on a line between Alice and Bob. Two neighboring referees share a singlet state, while $R_1$ and $R_N$ share singlet states with $A$ and $B$, resp.; each referee performs a joint measurement, leaving at the end the particles of Alice and Bob entangled. Whereas the quantum protocol has a straightforward and nice iterative character, we were not able to find a simulation protocol with a finite amount of communication in a $(N{+}1)$-locality scenario~\cite{branciard112}. Consider for instance the case with one additional referee $R_2$. Analogously to our Protocol~1, assume that Alice and $R_1$ share the random variable $\lambda_{AR_1}$, $R_1$ and $R_2$ share $\lambda_{R_1R_2}$, and $R_2$ and Bob share $\lambda_{R_2B}$, all uniformly and independently distributed on some interval $[0,\frac{\pi}{m}]$. After some finite communication, Bob could for instance (as in our first attempt, before Eq.~\eqref{eq_Prot0}) output $b=a=1$ if and only if $[\phi_A < \lambda_{AR_1}] = [\lambda_{AR_1} < \lambda_{R_1R_2}] = [\lambda_{R_1R_2} < \lambda_{R_2B}] = [\lambda_{R_2B} < \phi_B]$. This would result in the probability $P(a= b|\phi_A,\phi_B \in [0,\frac{\pi}{m}]) = \frac{m^3}{6\pi^3}|\phi_A-\phi_B|^3$, which scales cubically with $\phi_A-\phi_B$, and is therefore too small when $\phi_A$ is close to $\phi_B$. It is unclear how to change the cubic scaling with finite communication
. 
The following questions remain open: can multistage entanglement swapping be simulated with finite communication? Or can one prove, that above a certain value of $N$, an infinite amount of communication is necessary?

\paragraph{Acknowledgments.} We thank Jean-Daniel Bancal, Yeong-Cherng Liang, Stefano Pironio, Tim R\"az, and Ronald de Wolf for discussions. This work was supported by a UQ Postdoctoral Research Fellowship, the NSF grant CCF-0832787, the UK EPSRC, Canada's NSERC and CIFAR, the Swiss NCCR-QSIT, the US ARO, and the European ERC-AG QORE.

\bibliography{ent_swap_sim}

\begin{thebibliography}{10}
\bibitem{footnote_simul_BSM}%
  \BibitemOpen
  \bibinfo {note} {Note that we do not explicitly consider the simulation of
  the Referee's Bell state measurement (BSM). One could easily include this in
  the simulation protocol, by letting for instance the Referee and Bob share a
  random result for the BSM, and by letting Bob apply an approriate
  transformation to his measurement setting directly.}%
  \bibAnnoteFile{Stop}{footnote_simul_BSM}%
\end{thebibliography}

\section{Appendix}
\label{appendix}

In this appendix we prove that Protocols~1 and~2 generate the desired correlations~\eqref{corr}, or in other terms, $P(a=b|\bm{x},\bm{y}) = \frac{1 - \bm{x} \cdot \bm{y}}{2}$.

\subsection{Correlation from Protocol~1}

We first note that the definitions of $a = {\mathrm{sign}}(\sin \phi_A)$ and of $\phi_A' = (\phi_A{-}j_A\frac{\pi}{4} \mod \pi) \in [0,\frac{\pi}{4}[$ on Alice's side, and the definitions of $\beta = {\mathrm{sign}}[\sin (\phi_B-j_A\frac{\pi}{4})]$ (and then of $b = \pm \beta$) and of $\phi_B' = (\phi_B-j_A\frac{\pi}{4} \mod \pi) \in [0,\pi[$ on Bob's side, ensure that the following relations hold, as required:
\ba
P(a=b|\phi_A,\phi_B) &=& P(a\neq b|\phi_A,\phi_B+\pi) \nonumber \\
 && \hspace{-2cm} = P(a\neq b|\phi_A+\pi ,\phi_B) = P(a=b|\phi_A+\pi,\phi_B+\pi) \nonumber \\
 && \hspace{-2cm} = P(a=b|\phi_A{+}j\frac{\pi}{4},\phi_B{+}j\frac{\pi}{4}) \quad {\textrm{for \ any}} \ j \in \mathbb{Z}. \nonumber
\ea
It is therefore sufficient to check that the correct correlation is obtained for $\phi_A \in [0,\frac{\pi}{4}[$ and $\phi_B \in [0,\pi[$.

For such values of $\phi_A, \phi_B$ (for which $\phi_A' = \phi_A$ and $\phi_B' = \phi_B$), the probability $P(a=b|\phi_A,\phi_B)$ obtained from Protocol~1 can be calculated as follows:
\ba
&& P(a=b|\phi_A < \phi_B{-}j_B \frac{\pi}{4}) \nonumber \\
&& = \frac{16}{\pi^2} \, \Bigg( \int_{0}^{\phi_A} \!d\lambda_{AR} \int_{0}^{\lambda_{AR}} \!d\lambda_{RB} \ \wp_{001}^{j_B}(\phi_B{-}\lambda_{RB}) \nonumber \\
&& \qquad \quad + \int_{0}^{\phi_A} \!d\lambda_{AR} \int_{\lambda_{AR}}^{\phi_B{-}j_B\frac{\pi}{4}} \!d\lambda_{RB} \ \wp_{011}^{j_B}(\phi_B{-}\lambda_{RB}) \nonumber \\
&& \qquad \quad + \int_{0}^{\phi_A} \!d\lambda_{AR} \int_{\phi_B{-}j_B\frac{\pi}{4}}^{\frac{\pi}{4}} \!d\lambda_{RB} \ \wp_{010}^{j_B}(\phi_B{-}\lambda_{RB}) \nonumber \\
&& \qquad \quad + \int_{\phi_A}^{\phi_B{-}j_B\frac{\pi}{4}} \!d\lambda_{AR} \int_{0}^{\lambda_{AR}} \!d\lambda_{RB} \ \wp_{101}^{j_B}(\phi_B{-}\lambda_{RB}) \nonumber \\
&& \qquad \quad + \int_{\phi_A}^{\phi_B{-}j_B\frac{\pi}{4}} \!d\lambda_{AR} \int_{\lambda_{AR}}^{\phi_B{-}j_B\frac{\pi}{4}} \!d\lambda_{RB} \ \wp_{111}^{j_B}(\phi_B{-}\lambda_{RB}) \nonumber \\
&& \qquad \quad + \int_{\phi_A}^{\phi_B{-}j_B\frac{\pi}{4}} \!d\lambda_{AR} \int_{\phi_B{-}j_B\frac{\pi}{4}}^{\frac{\pi}{4}} \!d\lambda_{RB} \ \wp_{110}^{j_B}(\phi_B{-}\lambda_{RB}) \nonumber \\
&& \qquad \quad + \int_{\phi_B{-}j_B\frac{\pi}{4}}^{\frac{\pi}{4}} \!d\lambda_{AR} \int_{0}^{\phi_B{-}j_B\frac{\pi}{4}} \!d\lambda_{RB} \ \wp_{101}^{j_B}(\phi_B{-}\lambda_{RB}) \nonumber \\
&& \qquad \quad + \int_{\phi_B{-}j_B\frac{\pi}{4}}^{\frac{\pi}{4}} \!d\lambda_{AR} \int_{\phi_B{-}j_B\frac{\pi}{4}}^{\lambda_{AR}} \!d\lambda_{RB} \ \wp_{100}^{j_B}(\phi_B{-}\lambda_{RB}) \nonumber \\
&& \qquad \quad + \int_{\phi_B{-}j_B\frac{\pi}{4}}^{\frac{\pi}{4}} \!d\lambda_{AR} \int_{\lambda_{AR}}^{\frac{\pi}{4}} \!d\lambda_{RB} \ \wp_{110}^{j_B}(\phi_B{-}\lambda_{RB}) \Bigg) , \nonumber
\ea
and
\ba
&& P(a=b|\phi_A \geq \phi_B{-}j_B \frac{\pi}{4}) \nonumber \\
&& = \frac{16}{\pi^2} \, \Bigg( \int_{0}^{\phi_B{-}j_B\frac{\pi}{4}} \!d\lambda_{AR} \int_{0}^{\lambda_{AR}} \!d\lambda_{RB} \ \wp_{001}^{j_B}(\phi_B{-}\lambda_{RB}) \nonumber \\
&& \qquad \quad + \int_{0}^{\phi_B{-}j_B\frac{\pi}{4}} \!d\lambda_{AR} \int_{\lambda_{AR}}^{\phi_B{-}j_B\frac{\pi}{4}} \!d\lambda_{RB} \ \wp_{011}^{j_B}(\phi_B{-}\lambda_{RB}) \nonumber \\
&& \qquad \quad + \int_{0}^{\phi_B{-}j_B\frac{\pi}{4}} \!d\lambda_{AR} \int_{\phi_B{-}j_B\frac{\pi}{4}}^{\frac{\pi}{4}} \!d\lambda_{RB} \ \wp_{010}^{j_B}(\phi_B{-}\lambda_{RB}) \nonumber \\
&& \qquad \quad + \int_{\phi_B{-}j_B\frac{\pi}{4}}^{\phi_A} \!d\lambda_{AR} \int_{0}^{\phi_B{-}j_B\frac{\pi}{4}} \!d\lambda_{RB} \ \wp_{001}^{j_B}(\phi_B{-}\lambda_{RB}) \nonumber \\
&& \qquad \quad + \int_{\phi_B{-}j_B\frac{\pi}{4}}^{\phi_A} \!d\lambda_{AR} \int_{\phi_B{-}j_B\frac{\pi}{4}}^{\lambda_{AR}} \!d\lambda_{RB} \ \wp_{000}^{j_B}(\phi_B{-}\lambda_{RB}) \nonumber \\
&& \qquad \quad + \int_{\phi_B{-}j_B\frac{\pi}{4}}^{\phi_A} \!d\lambda_{AR} \int_{\lambda_{AR}}^{\frac{\pi}{4}} \!d\lambda_{RB} \ \wp_{010}^{j_B}(\phi_B{-}\lambda_{RB}) \nonumber \\
&& \qquad \quad + \int_{\phi_A}^{\frac{\pi}{4}} \!d\lambda_{AR} \int_{0}^{\phi_B{-}j_B\frac{\pi}{4}} \!d\lambda_{RB} \ \wp_{101}^{j_B}(\phi_B{-}\lambda_{RB}) \nonumber \\
&& \qquad \quad + \int_{\phi_A}^{\frac{\pi}{4}} \!d\lambda_{AR} \int_{\phi_B{-}j_B\frac{\pi}{4}}^{\lambda_{AR}} \!d\lambda_{RB} \ \wp_{100}^{j_B}(\phi_B{-}\lambda_{RB}) \nonumber \\
&& \qquad \quad + \int_{\phi_A}^{\frac{\pi}{4}} \!d\lambda_{AR} \int_{\lambda_{AR}}^{\frac{\pi}{4}} \!d\lambda_{RB} \ \wp_{110}^{j_B}(\phi_B{-}\lambda_{RB}) \Bigg) . \nonumber
\ea

One can then check that with the choice of functions $\wp_{c_A c_R c_B}^{j_B} \in [0,1]$ indicated in Table~\ref{table_ps}, this leads (for all values of $j_B$) to
\ba
P(a=b|\phi_A,\phi_B) &=& \frac{1 - \cos(\phi_A{-}\phi_B)}{2}, \nonumber
\ea
as desired.

\subsection{Correlation from Protocol~2}

After running Protocol~2 for inputs $\bm{x} =$ $(\sin{\theta_A}\cos{\phi_A}, \sin{\theta_A}\sin{\phi_A}, \cos{\theta_A})$ and $\bm{y} =$ $(\sin{\theta_B}\cos{\phi_B}, \sin{\theta_B}\sin{\phi_B}, \cos{\theta_B})$, the probability that Alice and Bob's outputs are the same is
\ba
&& P(a=b|\bm{x},\bm{y}) \nonumber \\
&& \quad = \sum_{a_0,b_0=\pm1} P(a_0,b_0|\phi_A,\phi_B) \, P(a=b|\theta_A,\theta_B,a_0,b_0) \nonumber \\
&& \quad = \sum_{a_0,b_0=\pm1} P(a_0,b_0|\phi_A,\phi_B) \, \frac{1 - \cos(a_0\theta_A{+}b_0\theta_B)}{2} \nonumber \\
&& \quad = \ P(a_0=b_0|\phi_A,\phi_B) \, \frac{1 - \cos(\theta_A{+}\theta_B)}{2} \nonumber \\
&& \quad \quad \quad + \ P(a_0\neq b_0|\phi_A,\phi_B) \, \frac{1 - \cos(\theta_A{-}\theta_B)}{2} \nonumber \\
&& \quad = \ \frac{1 - \cos(\phi_A{-}\phi_B)}{2} \, \frac{1 - \cos(\theta_A{+}\theta_B)}{2} \nonumber \\
&& \quad \quad \quad + \ \frac{1 + \cos(\phi_A{-}\phi_B)}{2} \, \frac{1 - \cos(\theta_A{-}\theta_B)}{2} \nonumber \\
&& \quad = \ \frac{1 - \cos\theta_A\cos\theta_B - \sin\theta_A\sin\theta_B\cos(\phi_A{-}\phi_B)}{2} \nonumber \\
&& \quad = \ \frac{1 - \bm{x} \cdot \bm{y}}{2} \,. \nonumber
\ea
Protocol~2 thus reproduces the desired entanglement swapping correlation~\eqref{corr}.

\end{document}